\documentclass{ws-procs9x6}
\usepackage{ulem}

      \def\new#1 {{\bf #1 }}
      \def\cut#1 {\sout{#1} }

\begin{document}

\title{Molecular gas in high redshift QSOs}

\author{C. L. Carilli$^1$, F. Bertoldi$^2$, F. Walter$^1$,
K.M. Menten$^2$, A. Beelen$^3$, P. Cox$^3$, A. Omont$^4$}
\address{~$^1$National Radio Astronomy Observatory\footnote{The
National Radio Astronomy Observatory is a facility of the National
Science Foundation operated under cooperative agreement by Associated
Universities, Inc.}, $^2$Max-Planck Inst. for Radio Astronomy,
$^3$Univ. Paris-Sud, $^4$Inst. d'Astrophysique de Paris \\ E-mail:
ccarilli@nrao.edu}


\maketitle

\abstracts{We review cm and mm observations of thermal molecular line
emission from high redshift QSOs. These observations reveal the
massive gas reservoirs (10$^{10}$ to 10$^{11}$ M$_\odot$) required to
fuel star formation at high rates. We discuss evidence for active star
formation in QSO host galaxies, and we show that these high
redshift, FIR-luminous QSOs follow the non-linear trend of increasing
L$_{FIR}$/L$'$(CO) with increasing L$_{FIR}$. We conclude with a brief
discussion of the recent CO detection of the most distant QSO at
$z=6.42$, and its implications for cosmic reionization.}

\section{Introduction}

Over the last few years, the study of high redshift QSOs has been
revolutionized in three ways. First, wide field surveys have revealed
100's of high z QSOs, right back to the epoch of cosmic reionization
($z > 6$; e.g., Fan et al. 2003). Second, it has been shown that most
(all?) low redshift spheroidal galaxies have central super-massive
black holes (SMBH), and that the black hole mass correlates with bulge
velocity dispersion. This M$_{BH}$-$\sigma_{\rm v}$ correlation
suggests coeval formation of galaxies and SMBH, thereby making SMBHs a
fundamental aspect of the galaxy formation process (Gebhardt et
al. 2000).  And third, mm surveys of high redshift QSOs find that
30$\%$ of the sources are `hyper-luminous infrared galaxies' ($L_{FIR}
= 10^{13}$ L$_\odot$), corresponding to thermal emission from warm
dust, and that this fraction is {\sl independent of redshift out to $z
= 6.4$} (Omont et al.  this vol.). If the dust is heated by star
formation, the implied star formation rates are extreme ($> 10^3$
M$_\odot$ year$^{-1}$), consistent with the formation of a large
elliptical galaxy on a dynamical timescale of 10$^8$ years.  On the
other hand, the FIR luminosity constitutes typically only 10$\%$ of
the bolometric luminosity of the sources, such that dust heating by
the AGN remains an alternative. Demographic studies show that SMBHs
acquire most of their mass during major accretion events marked by the
QSO phenomenon (Yu \& Tremain 2002).

Molecular line observations (typically CO) of FIR-luminous high $z$
QSOs have revealed large gas masses in most cases observed to date
(see Table 1).  Such gas reservoirs are a prerequisite for star
formation models for dust heating in FIR-luminous high $z$ QSOs. The
typical gas depletion timescales are of order 10$^7$ to 10$^8$ years,
if the dust is heated by star formation.  In this review we consider
this question in more detail.  We restrict ourselves to $z > 2$ QSOs
(see also Barvainis 1999; and see Scoville et al. 2004 and Sanders \&
Mirabel 1996 for observations of lower redshift QSOs). We assume a
standard concordance cosmology.

\section{Statistics}

Table 1 shows all the published detections of CO emission from sources
at $z > 2$ (only the most recent reference is given).  Most of the
sources are AGN (QSOs, radio galaxies = RG), since in these cases
optical spectroscopic redshifts are available, although a number of
optically brighter 'submm galaxies' (SMM) have now been detected in CO
emission (Neri et al. 2003). About half the sources are strongly
lensed (L). Most sources have been studied in the higher order
transitions ($\ge$ CO 3-2), although at $z \ge 4$ the lower order
transitions become accessible to cm telescopes such as the VLA.

\begin{table}
\tbl{CO sources at $z>2$ as of February 2004}
{\footnotesize
\begin{tabular}{lcccccl}
\hline
name & type & $z$ & Trans. &  L' & S$_{250}$ & ref \\
~    &  ~   & ~   & ~ & Jy km/s & mJy & ~ \\     
\hline
B1021+4724 & QSOL&  2.286 &  3-2 &   4.2  & 9.6 &  Solomon 1992 \\
J1636+4057 & SMM &  2.385 &  3-2 &   2.3  & 2.5 &  Neri 2003  \\
53W002     & QSO &  2.394 &  3-2 &   1.2  & 1.7 &  Alloin 2000 \\
J0443+0210 & SMM &  2.509 &  3-2 &   1.4  & 1.1 &  Neri  2003 \\
Cloverleaf & QSOL&  2.558 &  3-2 &   9.9  &  18 &  Weiss 2004 \\
J1401+0252 & SMML&  2.565 &  3-2 &   2.4  &  ~  &  Frayer 1999 \\
J1409+5628 & QSO &  2.583 &  3-2 &   4.0  & 10.7 & Beelen 2004 \\
B0414-0534 & QSOL&  2.639 &  3-2 &   2.6  &  40  & Barvainis  1998 \\
cB58       & SMML&  2.727 &  3-2 &   0.3  &  1   & Baker 2003 \\
J1230+1627 & QSOL&  2.735 &  3-2 &   0.80 &  2.7 & Guilloteau  1999 \\
J0239-0136 & SMML&  2.808 &  3-2 &   3.1  &  4.0 & Genzel 2003 \\
J2330+3927 & RG  &  3.094 &  4-3 &   1.3  &   ~  & de Breuck 2003 \\
B0751+2716 & QSOL&  3.200 &  4-3 &   6.0  &  6.7 & Barvainis 2002 \\
J0943+4700 & SMM &  3.346 &  4-3 &   1.1  &  2.3 & Neri 2003 \\
J0121+1320 & RG  &  3.520 &  4-3 &   1.2  &   ~  & de Breuck 2003 \\
J1909+722  & RG  &  3.534 &  4-3 &   1.6 &  4   & Papadopoulos 2000 \\
4C60.07    & RG  &  3.788 &  1-0 &   0.24 &  4.5 & Greve 2004 \\
B0827+5255 & QSOL&  3.911 &  1-0 &   0.15 & 17.0 & Papadopoulos 2001 \\
J2322+1944 & QSOL&  4.119 &  2-1 &   0.92 &  9.6 & Carilli 2003 Cox 2002 \\
B0952-0115 & QSOL&  4.434 &  5-4 &   0.91 &  2.8 & Guilloteau 1999 \\
B1335-0417 & QSO &  4.407 &  2-1 &   0.44 &  5.6 & Carilli 2002 \\
B1202-0725 & QSO &  4.693 &  2-1 &   0.49 & 12.6 & Carilli 2002 \\
J1148+5251 & QSO &  6.419 &  3-2 &   0.20 &   5.0 & Walter 2003
Bertoldi 2004 \\
\hline
\end{tabular}\label{table1} }
\vspace*{-13pt}
\end{table}

Figure 1 shows the correlation between FIR luminosity and velocity
integrated CO 1-0 luminosity ($L'_{CO(1-0)}$ K km s$^{-1}$ pc$^2$) for
sources at low and high redshift (Beelen et al. 2004).  The sources in
Table 1 consitute most of the sources with $L_{FIR} > 10^{13}$
L$_\odot$ in Figure 1. For sources without CO 1-0 measurements, the
1-0 luminosity was calculated assuming constant brightness
temperature.  For the high redshift sources the FIR luminosity is
given approximately by: $L_{FIR} = 4\times10^{12} (S_{250}/\rm mJy)
L_\odot$, appropriate for a typical ULIRG SED (Omont et al.  al. 2003;
note that in Fig 1 the values were calculated using the measured
multifrequency SEDs where available).  For dust heating by star
formation, the total star formation rate (SFR; from 0.1 to 100
M$_\odot$) is given by: $SFR = 4\times 10^{-10} L_{FIR}$ M$_\odot$
year$^{-1}$.  Gas masses can be derived from: M(H$_2$) = $X \times
L'_{CO(1-0)}$, where $X = 4$ for typical spiral galaxies, and $X =
0.8$ for ULIRGs (Downes \& Solomon 1998). Note that $X = 0.2$ is the
minimum (ie. optically thin) value, assuming solar C and O abundances,
and warm gas (70 K).  

\begin{figure}
\psfig{figure=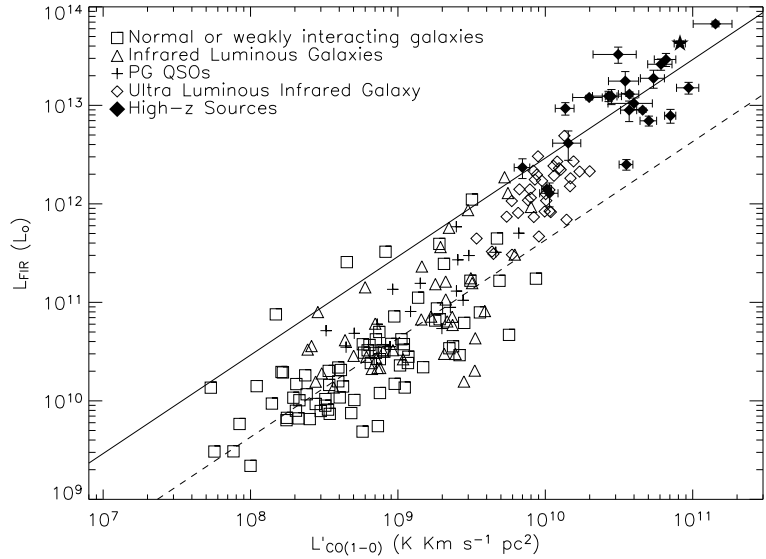,width=4.3in,angle=90}
\caption{The relationship between $L_{FIR}$ and $L'_{CO(1-0)}$
(Beelen et al. 2004).}
\end{figure}

The straight lines correspond to linear relations, with a ratio of
$L_{FIR}/L'_{CO} = 40$ (dash) or 300 (solid). The correlation is
clearly non-linear, and is better described by a power-law of the
form: $L_{FIR} \propto L'_{CO}$$^{1.7}$.  This non-linearity has been
interpreted as an increasing 'star formation efficiency' (= SFR/gas
mass) with increasing SFR (Gao \& Solomon 2003). An alternate
hypothesis is that the AGN dominates dust heating at the highest
luminosities. But this latter hypothesis begs the question: why would
there be any correlation at all? Meaning that if two very different
physical mechanisms are involved, one might expect a discontinuity in
this relationship going from starbursts to AGN.

\section{Examples}

Figure 2 shows examples of the CO excitation for three $z > 4$
QSOs. The line strengths are consistent with constant brightness
temperature (T$_B$) up to CO 5-4. This is similar to what is seen for
nuclear starburst regions, such as in M82, and implies dense ($> 10^4$
cm$^{-3}$), warm ($> 50$ K) gas. The recent detection of HCN emission
from the Cloverleaf quasar at $z =2.56$ supports the idea of dense,
warm (star forming?) molecular gas in these systems (Solomon et
al. 2003).

\begin{figure}
\psfig{figure=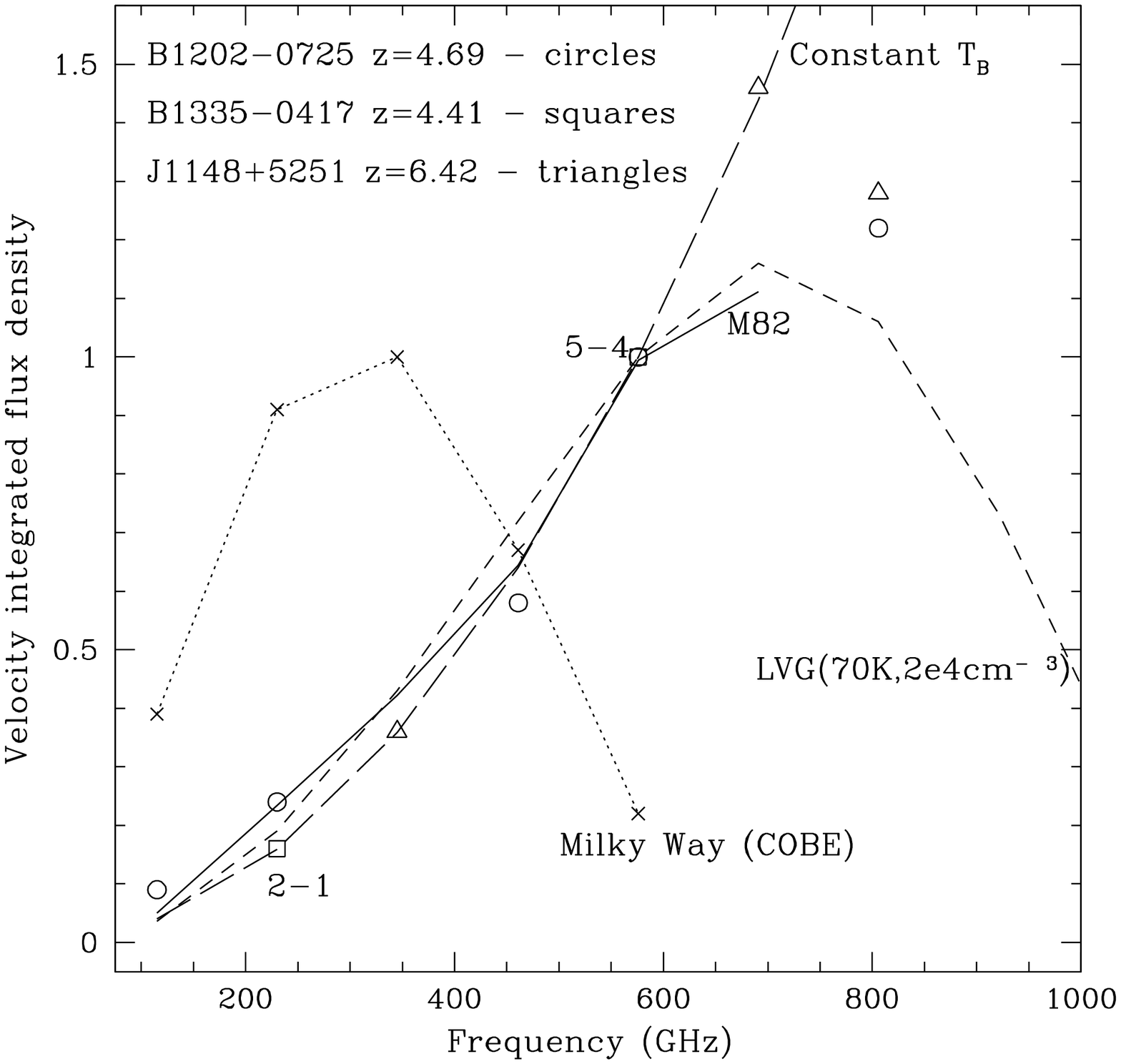,width=2.3in}
\vspace*{-2.3in}
\hspace*{2.3in}
\psfig{figure=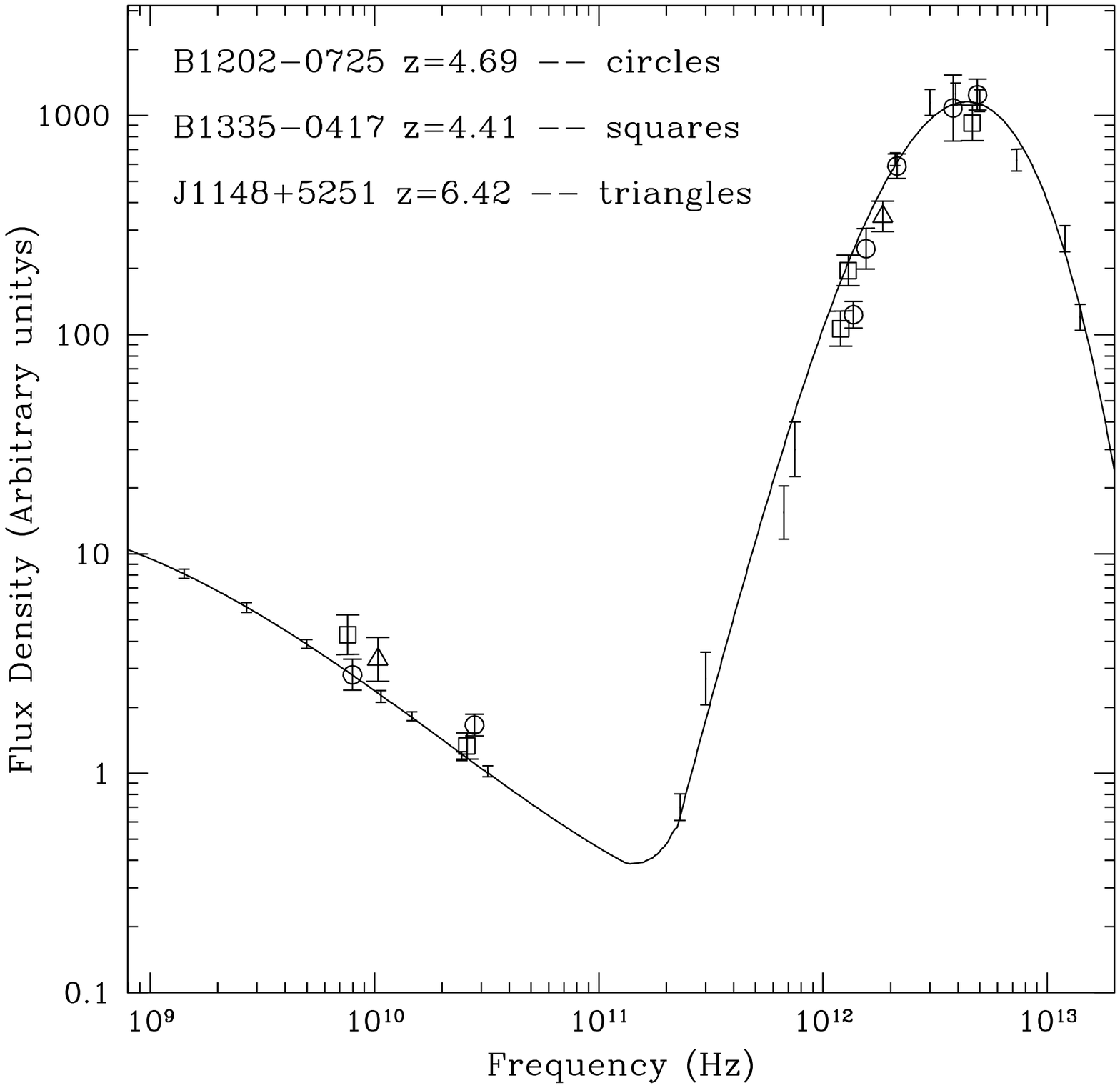,width=2.3in}
\caption{Left: The CO ladder for three high $z$ QSOs. 
Right: The radio to IR SEDs. The solid line is the
fit for M82.}
\end{figure}

Also shown in Figure 2 are the radio to IR SEDs of these
sources. Again, the SEDs are similar to those seen in typical star
forming galaxies.  VLBA imaging of the 1.4GHz continuum emission from
J1409+5628 at $z=2.58$ implies intrinsic brightness temperatures $<
10^5$ K, consistent with the non-thermal emission expected from a star
forming galaxy, but inconsistent with that expected from an AGN
(Beelen et al. 2004).

\begin{figure}
\psfig{figure=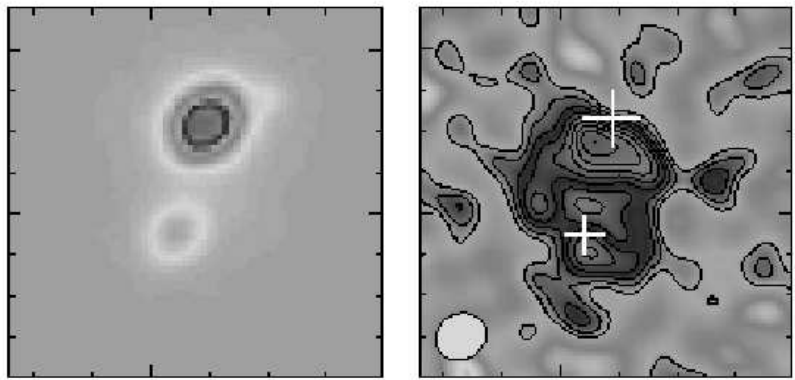,width=4.4in}
\caption{The optical (left) and CO 2-1 (right) images of the 
gravitationally lensed QSO J2322+1944 at $z=4.12$ 
(Carilli et al. 2003).}
\end{figure}

Figure 3 shows the optical and CO emission from the $z = 4.12$ QSO
2322+1944. The source is strongly gravitationally lensed, appearing as
a double QSO in the optical with 1$''$ separation (Djorgovski et
al. in prep), and a complete 'Einstein ring' in the CO 2-1 emission.
A similar ring is seen for the 1.4 GHz continuum emission (Carilli et
al. 2003). These results can be modeled in the source plane as a
starburst disk surrounding the QSO with a radius of about 2
kpc. J2322+1944 represents perhaps the best example of a coeval
starburst+AGN at high redshift.

We consider briefly the question of why only 30$\%$ of high $z$ QSOs
are FIR luminous?  Unfortunately, sensitivity limits of current
(sub)mm surveys cannot rule out a continuum of FIR luminosities
extending below a few mJy, although deep radio surveys suggest that
this may not be the case (Petric et al. 2004 in prep). If there really
are two types of QSOs (FIR-luminous and FIR-quiet), then the 30$\%$
fraction could represent a relative duty cycle for star formation
vs. QSO activity. Alternatively, there may be two types of major
accretion events for SMBHs: those that have associated star formation,
and those that do not.

\section{Cosmic Stromgren spheres}

The epoch of reionization (EoR) represents a fundamental benchmark in
cosmic structure formation, corresponding to ionization of the neutral
IGM by the first luminous sources. The recent discovery of Ly$\alpha$
absorption by the neutral IGM toward $z > 6$ QSOs (the Gunn-Peterson
effect; Fan et al. 2003) implies that these sources are situated at
the end of the EoR.

\begin{figure}
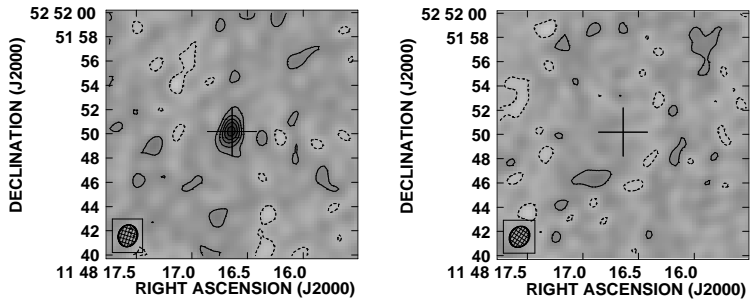

\psfig{figure=ON2.PS,width=2in}
\vspace*{-1.65in}
\hspace*{2in}
\psfig{figure=OFF3.PS,width=2in}
\vskip 0.1in
\caption{CO 3-2 emission from the highest redshift QSO,
J1148+5251 at $z=6.419$ (Walter et al. 2003). The left frame shows the
velocity integrated line emission at 1.5$''$ resolution at 46.7 GHz. The right
frame is an 'off' channel, showing the lack of continuum emission. The
contour levels are: -0.1,0.1,0.2,0.3,0.4,0.5 mJy/beam.} 
\end{figure}

Bertoldi et al. (2003) show that the 30$\%$ fraction of FIR-luminous
QSOs remains constant into the EoR, including the most distant QSO
known, J1148+5251, at $z=6.42$, which has an $L_{FIR} =
1.2\times10^{13}$ L$_\odot$.  CO emission has also been detected from
J1148+5251 using the VLA and the PdBI (Figure 4; Walter et al. 2003;
Bertoldi et al. 2003), with an implied molecular gas mass of $2\times
10^{10}$ M$_\odot$. The presence of a large mass of heavy elements and
dust in a galaxy just 0.8 Gyrs from the big bang raises interesting
issues for ISM enrichment, and suggests that active star formation
started at $z > 10$ in the host galaxy of this QSO. Also required is a
dust formation mechanism involving high mass stars and/or supernova
remnants.

A particularly interesting result for J1148+5251 is the difference
between the host galaxy redshift of 6.42 (as determined from the CO
line) and the on-set of the Gunn-Peterson absorption trough at
$z=6.32$. We note these CO observations are the only measurements
relating to the host galaxy properties (eg. the exact redshift), as
opposed to the AGN properties.  This redshift difference implies that
the QSO must be surrounded by a ionized sphere of physical radius
R=4.7 Mpc, presumably formed by the radiation from the QSO
itself. Hence, in J1148+5251 we are witnessing the process of
cosmic reionization. This is a 'time bounded' Stromgren sphere, and
implies a lifetime for the recent QSO activity of: $\rm t_{qso} = 10^7
(R/4.5Mpc)^3 f(HI)$ years, where f(HI) is the (volume averaged) IGM
neutral fraction (Walter et al. 2003; Wyithe \& Loeb 2003a). 

Wyithe \& Loeb (2003b) have used this result, plus models of QSO
formation, to argue (statistically) that the neutral fraction of the
IGM must be substantial at $z=6.4$: f(HI) $> 0.1$, otherwise the QSO
lifetimes become unreasonably short.  This neutral fraction is much
larger than the best current lower limits set by the GP effect of
f(HI) $> 0.001$, and argues for 'fast reionization' at $z \sim 6$
(Gnedin 2000). The implication is that f(HI) changes from $< 10^{-4}$
at $z = 5.7$ to $> 10^{-1}$ at $z = 6.4$, ie. the neutral fraction of
the universe changes by three orders of magnitude over a timescale of
only 0.1 Gyr.


\begin{thebibliography}{0}

\bibitem{gu} Alloin, D. et al. 2000, ApJ, 528, L81

\bibitem{gu} Baker, A. et al. 2004, astroph-

\bibitem{gu} Barvainis, R. 1999, in {\sl Highly Redshifted Radio
Lines}, (ASP: San Francisco), eds. Carilli et al. p. 39

\bibitem{gu} Barvainis, R. et al. 1998, ApJ, 492, L13

\bibitem{gu} Barvainis, R. et al. 2002, A\&A, 385, 399

\bibitem{gu} Beelen, A. et al. 2004, A\&A, submitted

\bibitem{gu} Beelen, A. et al. 2004, A\&A, in prep 

\bibitem{gu} Bertoldi, F. et al. 2003, A\& A, 409, L47

\bibitem{gu} Bertoldi, F. et al. 2003, A\& A, 406, L55

\bibitem{gu} Carilli, C. et al. 2003, Science, 300, 773

\bibitem{gu} Carilli, C. et al. 2002, AJ, 123, 1838

\bibitem{gu} Cox, P. et al. 2002, A\&A, 387, 406

\bibitem{gu} Downes, D. \& Solomon, P. 1998, ApJ, 507, 615

\bibitem{gu} de Breuck, C. et al. 2003, NewAR, 47, 285

\bibitem{gu} Fan, X. et al. 2003, AJ, 125, 1649

\bibitem{gu} Frayer, D. et al. 1999, ApJ, 514, L13

\bibitem{gu} Gao, Y. \& Solomon, P. 2003, ApJ (in press)
  astroph0310339 

\bibitem{gu} Gebhardt, K. et al. 2000, ApJ, 539, L13

\bibitem{gu} Genzel, R. et al. 2003, ApJ, 584, 633

\bibitem{gu} Gnedin, N. 2000, ApJ, 535, 530

\bibitem{gu} Greve, T., Ivison, R., Papadopoulos, P. 2004, A\&A, in press

\bibitem{gu} Guilloteau, S. et al. 1999, A\&A, 349, 363-368

\bibitem{gu} Neri, R. et al. 2003, 597, L113

\bibitem{gu} Omont, A. et al 2003, A\&A, 398, 857

\bibitem{gu} Papadopoulos, P. et al. 2001, Nature, 409, 58

\bibitem{gu} Papadopoulos, P. et al. 2000, ApJ, 528, 626

\bibitem{gu} Sanders, D. \& Mirabel, F. 1996, ARAA, 34, 749

\bibitem{gu} Scoville, N. et al. 2003, ApJ, 585, L105

\bibitem{gu} Solomon, P. et al. 2003, Nature, 426, 636

\bibitem{gu} Solomon, P., Downes, D., \& Radford, S. 1992 Nature, 356, 318

\bibitem{gu} Walter, F. et al. 2003, Nature, 424, 406

\bibitem{gu} Weiss, A. et al. A\&A,  409, L41

\bibitem{gu} Wyithe, S. \& Loeb, A. 2004a, astro-ph/0401188

\bibitem{gu} Wyithe, S. \& Loeb, A. 2004b, astro-ph/0401554

\bibitem{gu} Yu, Q. \& Tremaine, S. 2002, MNRAS, 335, 965

\end{thebibliography}
\end{document}